\documentstyle{article}
\def\ebn{\end{equation}\begin{equation}}
\begin{document}
\title{The Gauss Map and 2+1 Gravity}
\author{Raymond S. Puzio \\
\\
Department of Physics\\
Yale University\\
217 Prospect Street\\
New Haven, Conn, 06511\\
\\
e-mail: PUZIO\%YALPH2.bitnet\\
@yalevm.ycc.YALE.edu\\
}

\maketitle

\begin{abstract}
	We prove that the Gauss map of a surface of constant mean curvature
embedded in Minkowski space is harmonic.  This fact will then be used to
study 2+1 gravity for surfaces of genus higher than one.  By considering
the energy of the Gauss map, a canonical
transform between the ADM reduced variables and holonomy variables can
be constructed.  This allows one to solve (in principle) for the evolution
in the ADM variables without having to explicitly solve the constraints
first.
\end{abstract}
\maketitle

\section{Preamble}
	Given a surface embedded in Euclidean space, one defines the Gauss
map  in the following way:  at any point on the surface, one can construct
a (unique) normal vector of unit length.  A unit vector can be viewed as a
point on the sphere.  Thus we have a mapping from the embedded surface into
the sphere.  A result of Gauss states that, if the surface is minimal (and
hence has zero mean curvature), this
map is conformal.  A later result of Ruh and Vilms \cite{ruh} states that
if the map has constant (not necessarily zero) mean curvature, the Gauss
map is a harmonic map.  These results
are extremely useful in that they allow one to apply the vast body of
knowledge about conformal and harmonic maps between surfaces to the study of
constant mean curvature surfaces.  As an example of the power of these
results, witness the
formulas of Enneper, Weierstrass, and Kenmotsu which allow one to write the
parametric equation for a general surface of constant mean curvature as an
integral.

	To apply these results to general relativity in three dimensions, we
make the following two observations.  First, the theorems of Gauss, Ruh, and
Vilms apply to surfaces embedded in Minkowski space as well as to surfaces
in Euclidean space.  This will be proven explicitly later.  Second, in three
dimensions, the Einstein equations imply that spacetime is flat.  The reason
for this is that, in three dimensions (and only in three dimensions!), the
entire Riemann tensor is uniquely determined by the Einstein tensor.
Einstein's equations in vacuo imply the vanishing of the Einstein tensor, and
hence the vanishing of the entire Riemann tensor.  Therefore the resulting
spacetime is flat and locally isometric to Minkowski space, allowing us to
apply the aforementioned theorems to surfaces of constant mean curvature
embedded in these spacetimes.

	But why would anyone want to consider surfaces of constant mean
curvature embedded in a spacetime in the first place?  The reasons go back
to York's extrinsic time program and have nothing to do with Gauss maps
whatsoever.  In York's program, one considers slicing the spacetime into
spacelike slices of constant mean curvature and using the mean curvature
of the slice as time.  The reason York chose this slicing condition is that
it allows one to solve the Hamiltonian and momentum constraints of general
relativity separately instead of having to consider a nonlinear coupled
system of  partial differential equations.  Moreover, experience with actual
spacetimes has shown that constant mean curvature slicing is a good slicing in
the sense that constant mean curvature slices usually foliate the whole
spacetime, are unique, and do not develop singularities unrelated to the
singularities of the spacetime.  In contrast, Gaussian normal slicing, for
example, is a bad slicing since, even in Minkowski space, Gaussian normal
slices usually develop caustics after a finite lapse of time.

	York's extrinsic time prorposal can be made more exact and explicit
by Moincrief's reduction of the Einstein equations.  \cite{moncrief}
  In this approach, one
solves the constraints, substitutes the result back into the action, and
rearranges the result to obtain a Hamiltonian system with no constraints.
The configuaration space of the resulting system turns out to be
Teichm\"uller space, the space of equivalence classes of spatial metrics
under diffeomorphisms and conformal (Weyl) transforms.  Time is identified
with mean curvature, the momenta are transverse traceless tensor fields (long
known to mathematicians as a realization of the cotangent space to
Teichm\"uller space), and the Hamiltonian is equal to the area of the spatial
slice expressed as a function of the coordinates and their conjugate momenta.

	Considering this reduction approach reveals a remarkable fact about
the Gauss map.  The fact that the Gauss map is harmonic means that it only
depends on the conformal class of the spatial geometry.  Thus the Gauss map,
in a sense, automatically performs the reduction to Teichm\"uller space!
Contrast this with the usual situation where one has to put in the
Teichm\"uller parameters by hand by writing the metric as the product of a
conformal factor and a metric of constant negative curvature.  This fact is
all the more remarkable when one considers that a)  The unit hyperboloid in
Minkowski space has constant negative curvature.  b)  Eells, Earle, and
Sampson proved that one can represent Teichm\"uller space by metrics of
constant negative curvature by using harmonic maps!  To add to the amazement,
one finds that the energy of the Gauss map has a simple expression in terms
of dynamical quantities associated with the reduced Hamiltonian system.
These miraculous facts will be exploited extensively in a later section to
obtain information about the spacetimes.

\section{The Gauss Map in Minkowski Space}
	In this section we will prove the theorems of Gauss, Ruh, and Vilms
for spacelike surfaces in Minkowski space.  Since these theorems are local,
we can make life simple by using coordinate systems which can always be
defined in a small enough neighborhood of a given point but which do not
necessarily extend to coordinate systems on the whole of space and/or
spacetime.

	First we will demonstrate the result of Gauss that the Gauss map of
a surface of vanishing mean curvature is conformal.  To do so we first
introduce Cartesian coordinates on spacetime and isothermal coordinates
on space.  That is to say, we have spacetime coordinates t,x,y
\footnote{these coordinates will also be sometimes referred to as $x^0,
x^1$, and $x^2$.  In this article I will follow the convention that greek
indices denote spacetime indices $\alpha,\beta,\ldots=0,1,2$, latin
indices denote spatial indices $a,b,\ldots=1,2$, and indices inside
parentheses such as $e^{(\alpha)}$ denote frame indices}  such that
$ds^2 = -dt^2 + dx^2 + dy^2$ and spatial coordinates $\xi$ and $\eta$ such that
the induced spatial metric has the form $d\sigma^2 = e^{2 \lambda} (
d\xi^2 + d\eta^2)$.  The unit normal to the surface is given by taking the
(Minkowski space) cross product of two tangents to the surface and normalizing
to unit length:
\begin{equation}
n_{\alpha} =
{\epsilon_{\alpha \beta \gamma} x^{\beta}_{,\xi} x^{\gamma}_{,\eta}
\over \sqrt{x^{\alpha}_{,\xi} x_{\alpha ,\xi}
x^{\beta}_{,\eta} x_{\beta ,\eta} -
(x^{\alpha}_{,\xi} x_{\alpha ,\eta})^2 } }
\end{equation}
The fact that the metric $e^{2 \lambda}(d\xi^2 + d\eta^2)$ is induced from
the Minkowski metric implies the following two equations:
\begin{equation}
x^\alpha_{,\xi} x_{\alpha,\xi} = x^\alpha_{,\eta} x_{\alpha,\eta}
\end{equation} \begin{equation}
x^\alpha_{,\xi} x_{\alpha,\eta} = 0
\end{equation}
Using the above two equations allows us to simplify the denominator in the
equation for the unit normal:
\begin{equation}
n_\alpha = {\epsilon_{\alpha \beta \gamma} x^\beta_{,\xi} x^\gamma_{,\eta}
\over x^\beta_{,\xi} x_{\beta,\xi}} =
{\epsilon_{\alpha \beta \gamma} x^\beta_{,\xi} x^\gamma_{,\eta}
\over x^\beta_{,\eta} x_{\beta,\eta}}
\end{equation}
Now consider the quantity $H^\alpha := (\partial_{\xi \xi} +
\partial_{\eta \eta}) x^\alpha$.  By differentiating equations (2) and (3)
with respect to $\xi$ and $\eta$, we see that $H^\alpha x_{\alpha,\xi} =
H^\alpha_{\alpha,\eta} = 0$  Thus $H^\alpha$ is normal to the surface.  I
claim that the length of $H^\alpha$ is proporional to the mean curvature.
To see this, let
us Lorentz transform our coordinates such that $x=\xi$ and $y=\eta$ to
first order at some given point P.  Then it is easy to compute the mean
curvature at P by differentiating equation (4):
\begin{equation}
\tau = g^{ab}n_{a;b} = e^{-2\lambda} (n_{x,\xi} + n_{y,\eta}) =
e^{-2\lambda}(-t_{,\xi\xi}-t_{,\eta\eta}) = -e^{-2 \lambda} H^0
\end{equation}
Thus $H^\alpha = -e^{2\lambda} \tau n^{\alpha}$.   In particular, if our
surface has vanishing mean curvature, $H^\alpha = 0$

	To show that vanishing mean curvature implies conformality of the
Gauss map, introduce a complex parameter $z := \xi + i \eta$ and the
quantities $w^\alpha := x^\alpha_{,z}$.  The pair of real equations (2) and
(3) can be combined into the complex equation
\begin{equation}
g_{\alpha\beta} w^\alpha w^\beta = 0
\end{equation}
This complex equation can be seen as defining a curve in 3-dimensional
complex projective space which is conformal to the unit hyperboloid in
Minkowski space.  With this identification, $w^\alpha(z)$ is the Gauss
map.  In this complex notation, $H^\alpha = x^\alpha_{,z\overline{z}} =
w^\alpha_{,\overline{z}}$.  Thus, when our surface has vanishing mean
curvature, $w^\alpha_{,\overline{z}} = 0$, which means that the Gauss
map is holomorphic, hence conformal. Q.E.D.

\proclaim Theorem. The Gauss map of a spacelike surface embedded in
Minkowski space is conformal

	Now we will consider a surface of constant mean curvature
embedded in Minkowski space and prove the theorem of Ruh and Vilms which
states that its Gauss map is harmonic.  Whereas in the previous section
Cartesian coordinates proved convenient, Gaussian normal coordinates
will be useful in this section.  Let $\gamma_{ab}$ be the spatial
metric (Unlike last section, the exact form of $\gamma_{ab}$ is irrelavant
in this section).  Then the metric on spacetime is given by $-dt^2 +
\gamma_{ab} dx^a dx^b$ and the surface of constant mean curvature is given
by setting t=0.  Introduce an orthonormal frame of paralell vectors
$e^{(\alpha)}_\beta$.  By paralell I mean that $e^{(\alpha)}_{\beta;\gamma}
= 0$ ,i.e. that the frame would be the coordinate frame for Cartesian
coordinates.  By definition of normal coordinates, the unit normal
vector is $n^\alpha = (1,0,0)$ and hence the Gauss map is given by
$n^{(\alpha)}(x,y) := e^{(\alpha)\beta}n_\beta (x,y)$.

	To show that the gauss map is harmonic, we will directly compute
the variation of the energy and show that it is stationary for the gauss
map.  Putting the usual metric on the unit hyperboloid, we write down the
harmonic map energy for the Gauss map as
\begin{equation}
E = \int \eta_{\alpha\beta} n^{(\alpha)}_{,a} n^{(\beta)}_{,b}
\gamma^{ab} \gamma^{-1/2} d^2 x
\end{equation}
where $\eta_{\alpha \beta} := diag(-1,1,1)$.
The variation of the energy is given by
\footnote {Very soon we will apply the theorem being proven here to the case
where the spacetime is only locally Minkowski.  However, this does not
invalidate the current proof since, in varying an action, it is sufficient to
consider variations $\delta n^{(\alpha)}$ which are supported inside an
arbitrarily small neighborhood.  The neighborhood can be picked small enough
that the coordinates and frame used here can be defined.}
\begin{equation}
\delta E = 2 \int \eta_{\alpha\beta} \delta n^{(\alpha)}_{,a} n^{(\beta)}_{,b}
\gamma^{ab} \gamma^{-1/2} d^2 x
\end{equation}
Now, the quantity $\delta n^{(\alpha)}$ is only defined on the surface and we
are free to define it off of the slice in any consistent way.  For simplicity,
we will define it by setting ${\partial \delta n^{(\alpha)} \over \partial t}
= 0$.  Then we will not change the value of $\delta E$ if we replace the
ordinary derivatives by covariant derivatives and $\gamma$ by g.
\begin{equation}
\delta E = 2 \int \eta_{\alpha\beta} \delta n^{(\alpha)}_{;\gamma}
n^{(\beta)}_{;\delta}
g^{\gamma \delta} g^{-1/2} d^2 x
\end{equation}
Now we may integrate by parts
\begin{equation}
\delta E = -2 \int \eta_{\alpha\beta} \delta n^{(\alpha)}
n^{(\beta)}_{;\gamma \delta}
g^{\gamma \delta} \gamma^{-1/2} d^2 x
\end{equation}
Now we note that $n_{\alpha;0} = n_{0;\alpha} = 0$ in Gaussian normal
coordinates, hence we obtain
\begin{equation}
n^{(\beta)}_{;\gamma \delta} =
e^{(\beta) \epsilon} n_{\epsilon;\gamma \delta} =
e^{(\beta) \epsilon} n_{\gamma;\delta \epsilon}
\end{equation} \begin{equation}
g^{\gamma \delta} n^{(\beta)}_{;\gamma \delta} =
e^{(\beta) \epsilon} (g^{\gamma \delta} n_{\gamma;\delta})_{;\epsilon} =
e^{(\beta) \epsilon} (\gamma^{ab} n_{a;b})_{;\epsilon}
\end{equation}
Substituting into equation (10),
\begin{equation}
\delta E = -2 \int \eta_{\alpha\beta} \delta n^{(\alpha)}
e^{(\beta) \epsilon} (\gamma^{ab} n_{a;b})_{;\epsilon}
g^{-1/2} d^2 x
\end{equation}
Remembering that ${\partial \delta n^{(\alpha)} \over \partial t}
= 0$, we may finally simplify the integral to
\begin{equation}
\delta E = -2 \int \eta_{\alpha\beta} \delta n^{(\alpha)}
e^{(\beta) a} \tau_{,a}
g^{-1/2} d^2 x
\end{equation}
Thus, if $\tau$ is constant on the surface, $\delta E = 0$, and hence the
Gauss map is harmonic. Q.E.D.

\proclaim Theorem. The Gauss map of a spacelike surface of constant mean
curvature embedded in Minkowski space is harmonic.

\section{Relation to the Reduced Hamiltonian System}

	In this section we will relate the different objects which appeared
in the Gauss map to objects appearing in the Hamiltonian reduction.  First,
by considering the holonomies of our spacetime, we will show that the
target space for the gauss map can be thought of as a closed Riemann surface
whose Teichm\"uller parameters are given by certain constants of the motion.
This will allow us to apply various theorems on harmonic maps between
Riemann surfaces.  Next we will compute the energy of the harmonic map and
show how it may be related to the area of the slice, which plays the role of
Hamiltonian in the reduced formalism.  In all cases, the genus of the
spatial manifold is assumed to be higher than 1.

	First, we consider the holonomies of the spacetime.  Since the
spacetime is flat, parallel transport along curves which are continuously
deformable into each other will be independant of the choice of curve.  Thus,
given an element $[c] \in \pi_1(P)$, i.e. an equivalence class of
non-contractible loops passing through the point P, we can define the
holonomy associated with [c] to be the matrix $H^\alpha_\beta([c])$ which
describes parallel transport around a loop in [c].  Given a vector $v^\mu$
at the point P, the vector goes into the vector $H_\nu^\mu([c])v^\nu$
upon parallel translation about any loop in [c].  The holonomies form a
representation of the fundamental group in the natural way:  given two curves
$[c_1]$ and $[c_2]$, the holonomy associated with $[c_1 \circ c_2]$ is
$H^\mu_\nu([c_1])H^\nu_\xi([c_2])$

	To relate this to the Gauss map, consider the action of the
holonomies on the normal vector.  Thinking of the unit vector as point on the
unit hyperboloid, the effect of the holonomy is a motion of the hyperboloid.
In the usual negative curvature metric on the hyperboloid, these motions
are isometries.  Discrete groups of such motions have been studied
extensively under the names of Fuchsian groups and the reader is referred to
\cite{katok} for details.  All that we need to know for the present is that
we can quotient the unit hyperboloid by the group of holonomies and that
this results in a nonsingular surface of constant negative curvature of the
same genus as our original surface as long as the holonomy group is not
degenerate.  As we shall see when we consider holonomies in more detail,
nondegeneracy will generically be true, so assuming nondegeneracy is
justified.

	As a map from the slice of constant mean curvature to the unit
hyperboloid, the gauss map is multiple-valued because the unit normal
changes by a holonomy matrix every time one traverses a closed loop.
However, if one instead considers the Gauss map as a map into the surface
gotten by quotienting the hyperboloid as above, then the resulting map
is single-valued; the holonomies and the Fuchsian matrixes cancel out.
Since harmonicity is a local property, the new Gauss map into this
surface is also a harmonic map.

	The conformal class of the target surface does not change with time.
To see this, first note that the trace of a holonomy is invariant under
changing the base point on the loop, and hence, under arbitrary smooth
deformations of the loop.  In particular, we may push the loop forward in
time, hence the trace is a constant of the motion.  Second, the Fuchsian
group giving our surface can be determined uniquely up to cojugation by
specifying the traces of its elements.  In fact, counting degrees of
freedom shows that it is enough to give 6g-6 traces to specify the group
for a genus g surface.

	Thus, evolution may be viewed as the evolution of a harmonic map
between two Riemann surfaces of the same genus.  To understand this evolution
better, we can make use of the many theorems on harmonic maps between Riemann
surfaces.  For the present we will consider three theorems.  For discussion
and proofs refer to \cite{jost}

\proclaim Theorem 1. For any map $f: \Sigma_1 \rightarrow \Sigma_2$,
$Area(f(\Sigma_1)) \leq E(f)$ with equality iff f is conformal.

\proclaim Theorem 2. Suppose that $\Sigma_1$ and $\Sigma_2$
are compact surfaces without
boundary, and that $h: \Sigma_1 \rightarrow \Sigma_2$ is a diffeomorphism.
Then there exists a harmonic diffeomorphism $u: \Sigma_1 \rightarrow \Sigma_2$
isotopic to h.  Furthermore, u is of least energy among all diffeomorphisms
isotopic to h.

\proclaim Theorem 3.  Let $\Sigma_1$ and $\Sigma_2$ be Riemann surfaces of
the same genus and let $R(\Sigma_2) < 0$.   Then any map of degree one is
uniquely homotopic to a harmonic map which is a diffeoemorphism.

	These theorems show us that our map is uniquely determined by the
geometry of the slice and that it is well-behaved.  Since this map is unique
and the target space is fixed, the energy of this map is a well-defined
function of the geometry on the source space.  It is obvious from the
definition that E cannot depend on the conformal factor in the geometry on
the source space.  Moreover, if one makes a diffeomorphism of the source
geometry ${\gamma'}_{ab}(y(x)) = \gamma_{cd}(x) {\partial y^c \over
\partial x^a} {\partial y^d \over \partial x^b}$, the transformed map
${u'}^{(\mu)}(y) = u^{(\mu)}(y(x))$ will be the unique solution to the harmonic
map equation with $\gamma'_{ab}$ and will have the same energgy as u.
Therefore E can be considered as a function of the Teichm\"uller parameters
of the source space for a fixed target space.

	Seeing that E can be thought of as a function of the Teichm\"uller
parameters of the spatial slice and the target space metric which is
parameterized by the holonomies, one wonders whether the value of E has any
interpretation.  To answer this question, let us calculate.  As previously,
introduce Gaussian normal coordinates:
\begin{equation}
E  = \int \eta_{\mu\nu} n^{(\mu)}_{,a} n^{(\nu)}_{,b} \gamma^{ab}
\gamma^{1/2} d^2 x \ebn
 =  \int \eta_{\mu\nu} n^{(\mu)}_{;a} n^{(\nu)}_{;b} g^{ab}
g^{1/2} d^2 x \ebn
 =  \int \eta_{\mu\nu} e^{(\mu)c} n_{c;a} e^{(\nu)d} n_{d;b} g^{ab}
g^{1/2} d^2 x \ebn
 =  \int g^{cd} n_{c;a} n_{d;b} g^{ab} g^{1/2} d^2 x \ebn
 =  \int \gamma^{ab} \gamma^{cd} K_{ac} K_{bd} \gamma^{1/2} d^2 x
\end{equation}
Now recall the definitions of the gravitational momentum and the Hamiltonian
constraint:
\begin{equation}
\pi^{ab} = \gamma^{1/2} (\gamma^{ab} K^c_c - K^{ab})
\ebn
\gamma^{-1/2} (\pi^{ab}\pi_{ab} - (\pi^a_a)^2) = \gamma^{1/2} R
\end{equation}
Combining and integrating we obtain
\begin{equation}
\gamma^{1/2} K^{ab} K_{ab}  = \gamma^{1/2} ((K^c_c)^2 + R)
\ebn
 E = A \tau^2 + 4 \pi \chi
\end{equation}
Thus one sees that the energy can be written in terms of the area of the
surface and the mean curvature.  When one considers that the area is the
reduced Hamiltonian, the above formula gains in significance.  To properly
appreciate it, we must first consider some facts about canonical transforms
and constants of the motion.

\section{Canonical Transforms and Constants of the Motion}

	In the last section, we considered the holonomies associated with
parallel transport.  As it turns out, there is another way to introduce
holonomes into 2+1 gravity.  This approach, which was found by Witten,
\cite{witten}
give holonomies as path ordered integrals of matrices constructed from the
frame and spin connection.  Like the holonomies considered above, they are
constants of the motion.  In fact they provide a {\em complete} set of
constants of the motion, allowing one to unambiguously specify any 2+1
spacetime by its holonomies.

	Let us recall the definitions and properties of these holonomies.
Pick an orthonormal frame $e^{(\alpha)}_\beta$ for the spacetime and construct
its spin connection $\omega^{(\alpha)}_{(\beta) \mu}$.  Then the holonomy
associated with the loop C is:
\begin{equation}
w_r (C) = tr~P~exp (\int \omega^{(\alpha)}_{(\beta)\mu} J_\alpha^\beta
+ e^{(\alpha)}_\mu P_\alpha dx^\mu)
\end{equation}
The subscript ``r'' in $w_r$ refers to a representation of ISO(2,1) and
J refers to the generator of boosts/rotations in that representation and
P refers to the generator of translations.  Since the space of solutions
for relativity in 2+1 dimensions is 12g+12 dimensional, not all of these
holonomies can be independant functions.  In fact it is enough to consider
just two representations and 6g-6 loops.  Two representations which do the
job are given below.

	In the first representation, J and P are given by:
\begin{equation}
J^0_1 = 1/2 \pmatrix{0&1 \cr 1&0}  \qquad
J^1_2 = 1/2 \pmatrix{i&0 \cr 0&-i} \qquad
J^2_0 = 1/2 \pmatrix{0&-i \cr i&0} \qquad
\ebn
P_1 = P_2 = P_3 = 0
\end{equation}
In this representation, tr denotes the usual sum of the two diagonal
elements of the matrix.

	The second representation is 4 dimensional.  The matrices that
represent J and P are given in block form by
\begin{equation}
J^\alpha_\beta = \pmatrix {j^\alpha_\beta &0 \cr 0&j^\alpha_\beta} \qquad
P_\alpha = \epsilon_{\alpha\beta\gamma} \pmatrix{0&j^\beta_\gamma \cr 0&0}
\end{equation}
$j^\alpha_\beta$ denotes the matrixes $J^\alpha_\beta$ in the first
representation, i.e. $j_\beta^{\alpha \quad this \quad paragraph}$ =
$J_\beta^{\alpha \quad last \quad paragraph}$.  \footnote{
Essentially this representation is the same as
the representation given in \cite{martin}, but without the funny
infinitessimals which square to zero.} In this representation traces are
taken by the rule $tr M = M_{31} + M_{42} + M_{13} + M_{24}$.  From now on
$w_1(C)$ will refer to the holonomy in the first representation and $w_2(C)$
will refer to that in the second.
	It turns out that the holonomies in the first representation,
$w_1(C)$, equal the holonomies employed in the last section.  To see this,
recall the result of Waelbroeck \cite{waelbroeck} that any 2+1 vaccum
spacetime can be realized
as the quotient of Minkowski space by an appropriate discrete subgroup of
ISO(2,1).  Then, given a closed curve $C$ in the spacetime, it can be
considered as coming from an open curve $\tilde C$ in Minkowski space whose
endpoints are identified by the action of an elememt $\cal T$ of
ISO(2,1).  Now, it is clear that the holonomy of $C$ in the sense of the last
section is simply the boost part of $\cal T$.  To compute the holonomy in
the sense of Witten, let us assume that $\tilde C$ is a helicoid. (We lose no
generality since the holonomies are invariant under deformations.)  Namely,
let us write ${\cal T} = e^{p+j}$ where $j$ is a boost generator and $p$ is a
translation generator.  Then we assume that $C(s) = e^{(p+j)s}P_0$ where
$P_0$ is the initial endpoint of $C$.  To compute $w_1(C)$, we introduce a
framing in the following way: Consider the plane through P normal to $C$.
Introduce vectors $e^{(0)}_a$ and $e^{(1)}_a$ which form a Cartesian frame
for this plane.  Let the vector $e^{(2)}_a$ be the normal to this plane.
Any point in spacetime not on the plane can be mapped into a point on the
plane by a matrix $e^{(j+t)\tilde t}$ for some suitable $\tilde t$.  Extend
the frame off of the surface by using the action of the matrix
$e^{(j+t)\tilde t}$.  Then we have a frame in which $e^{(2)}_a$ is tangent to
the curve $C$ and the other two vectors are orthogonal to it.  Computing the
connection is quite easy if we introduce Cartesian coordinates in the
spacetime, for then  we have $\omega^{(\alpha)}_{(\beta) \mu} = e^\nu_{(\beta)}
e^{(\alpha)}_{\nu,\mu}$.  For the purpose of computing the holonomies, it
is enough to know the connection along the curve for a direction tangent to
the curve.  If we pick our cartesian coordinates at some point $P$ on the
curve such that $e^{(\alpha)}_\beta = \delta^\alpha_\beta$ at P, then we
can readily see that $\omega^{(\alpha)}_{(\beta)2} J^\beta_\alpha = m$.  This
means that $w_1(C) = tr~P~e^{\smallint m} = tr~e^m$.  Thus the two holonomies
agree.

	Having shown that the holonomies we considered in the last section
are the same as Witten's holonomies, we are now in a position to use the
results about those holonomies given by Martin \cite{martin}.
According to Martin, the Poisson brackets
of two $w_1$'s is zero.  Let us denote the Teichm\"uller parameters of
the unit hyperboloid quotiented by the holonomies by $Q^A$. (Capital latin
indices will run form $1 \ldots 6g-6$ and indicate quantilties tht have
something to do with Teichm\"uller space.) The $Q^A$'s can be expressed as
functions of the $w_1$'s, and hence, Poisson commute.  Since the $w_1$'s
and the $w_2$'s together form a complete set of observables, we can form
functions $P_A(w_1,w_2)$ which are canonically conjugate to the $Q^A$'s.

	Thus we have two sets of canonical variables on our phase space: the
$q^A$ and $p_A$ coming from the Hamiltonian reduction and the $Q^A$ and
$P_A$ coming form the holonomies.  Moreover, the $Q^A$ and $P_A$ are
constants of the motion.  This is a standard situation in Hamiltonian
mechanics and is described by a (time-dependant) generating function.
Recall that, if $h$ is the hamiltonian in some canonical variables $q^i,p_i$,
$H$ is the Hamiltonian in terms of some other canonical variables $Q^i,P_i$,
and the transform between the two sets of variables is governed by a
generating function $F(q^i,Q^i,t)$, we then have the equations:
\begin{equation}
p_i = {\partial F \over \partial q^i} \qquad
P_i = {\partial F \over \partial Q^i}
\ebn
H(q^i,Q^i,t) = h(q^i,Q^i,t) + {\partial F \over \partial t} (q^i,Q^i,t)
\end{equation}
In our case, H is zero since the $Q^i$ and $P_i$ are constants of the motion
and h is equal to the area of the constant mean curvature slice.  We need
h expressed in terms of $q^i$ and $Q^i$.  But this exactly what we obtained
in equation (23) in terms of the energy of the Gauss map! Thus we have
\begin{equation}
h(q^i,Q^i) = A(q^i,Q^i) = \tau^{-2} (E(q^i,Q^i) - 4 \pi \chi)
\end{equation}
Comparison with equation (29) allows us to read off the generating function F
\begin{equation}
F(q^i,Q^i) = A(q^i,Q^i) = {1 \over 2 \tau^3} (E(q^i,Q^i) - 4 \pi \chi)
\end{equation}
which allows us to solve (in principle) for the relation between the ADM
reduced variables and the holonomies.

\section{Solving for E(q,Q) in Principle}

Having made the claim that one can solve for the relation between q and Q in
principle several times, it is time to make good on that claim by showing a
method that works in principle although appears intractable in practise.  To
begin, we need the harmonic map whose energy we are to obtain.  To do so we
could make use of the following fact:  If $f:M \rightarrow N$ is a harmonic
map and the image of $f$ lies on a submanifold $N_1 \subset N$ (with induced
metric), then $f:M \rightarrow N_1$ is a harmonic map.  In our case $N$ is
Minkowski space and $N_1$ is the unit hyperboloid.  For a function into
Minkowski space to be as harmonic map means that each of the three coordinate
functions satisfies the Laplace equation, i.e. be harmonic in the old sense of
the term.  For a function on a surface (in this case our spatial slice) to be
harmonic is equivalent to demanding that it be the real part of an analytic
function.  Thus we are led to consider the automorphic functions on our
surface.  In particular, we want to find three such functions $\alpha,\beta,
\gamma$ such that 1) Their real parts lie on the unit hyperbolid 2) They are
multiple valued and, when one goes around some non-contractible loop on the
double torus, the corresponding point on the hyperboloid is the one coming
from acting on the original point with an element of $\Gamma_Q$.  Since
automorphic functions, like the elliptic functions they generalize, are
determined by their singularities,  we can arrange these conditions by looking
at what happens at the singularities(which will be branch points).  Then to
find the energy one would need to integrate the harmonic map energy over the
surface.

There is nothing impossible about the above plan; it is just tedious.  Since
the final result would be a complicated surface integral which almolst
certainly could not be done in closed form, it is doubtful how much of the
plan is worth carrying out.  However it strongly indicates one important
thing: the reduced constant-mean-curvature Hamiltonian is certainly not as
nice an object in the higher-genus case as on the torus.  It probably cannot
be written in closed form and is not simply the Weil-Peterson line element as
one might have hoped from simple analogy with the torus.

\section{Conclusions}

	We have succeded in obtaining the reduced Hamiltonian for constant mean
curvature slicing without solving the constraints.  The only catch is that
it is given in terms of the function $E(q^A,Q^A)$.  This is some complicated
function and, presumably, the only way to get at it is by solving for the
unique harmonic map between two Riemann surfaces.  Thus, from a practical
point of view, all we did was replace one quasilinear elliptic equation (the
Lichnerowitz equation) with another quasilinear elliptic equation (the
harmonic map equation).  As far as getting an exact solution to the problem,
we seem to have gained little or nothing.

	However, I consider the real value of equation (31) to be the way in
which it unites three subjects in a natural way.  The quasilinear elliptic
equation we are obliged to consider is not just any old nonlinear equation,
but one that has been studied extensively over the past three decades by both
mathematicians (as the harmonic map equation) and physicists (as the nonlinear
$\sigma$-model).  Thus it allows one to apply results about that subject to
the study of 2+1 gravity.  Also, the holonomies of Witten have appeared in
this study and been related to the ADM variables in a canonical way.

Even if the function $E(q,Q)$ cannot be gotten at in any useful form, one might
be able to say something about its limiting values.  One might, for example,
be able to compute it asymptotically in the limiting case where the
Teichm\"uller parameters
$q$ are describing a double torus pinching off into a pair of single tori.
Then, one could make concrete calculations regarding this kind of topology
change.

	In the conclusion to his paper \cite{moncrief}, Moncrief said that
the solution to the higher genus case looked rather remote.  It still seems
so.  However, this Gauss map construction brings us slightly closer and
allows us to glimpse some general features of that solution.

\section{Acknowledgements}

	I thank Sanjaye Ramgoolam and Vincent Moncrief for useful discussions
and advice.

\end{document}